\documentclass[12pt]{article}

\usepackage{amsmath,amssymb,graphicx} 
\usepackage{epsf}

\newcommand{\arline}{\nonumber \\}
\newcommand{\centeron}[2]{{\setbox0=\hbox{#1}\setbox1=\hbox{#2}\ifdim
                           \wd1>\wd0\kern.5\wd1\kern-.5\wd0\fi \copy0
                           \kern-.5\wd0\kern-.5\wd1\copy1\ifdim\wd0>\wd1
                           \kern.5\wd0\kern-.5\wd1\fi}}
\newcommand{\ltap}{\>\centeron{\raise.35ex\hbox{$<$}}
                   {\lower.65ex\hbox{$\sim$}}\>}
\newcommand{\gtap}{\>\centeron{\raise.35ex\hbox{$>$}}
                   {\lower.65ex\hbox{$\sim$}}\>}
\newcommand{\gsim}{\mathrel{\gtap}}
\newcommand{\lsim}{\mathrel{\ltap}}

\newcommand\ZZ{\hbox{\zfont Z\kern-.4emZ}}
\font\zfont = cmss10 

\renewcommand{\thefootnote}{\fnsymbol{footnote}}

\def\o{\omega}

\def\eps{\epsilon}
\def\tSigma{\tilde{\Sigma}}
\def\CR{\nonumber \\ }
\def\leqn#1{(\ref{#1})}
\newcommand\iden{\leavevmode\hbox{\small1\normalsize\kern-.33em1}}
\def\draftnote#1{{\bf #1}}
\def\beqa{\begin{eqnarray}}
\def\eeqa#1{\label{#1}\end{eqnarray}}
\def\eeqan{\end{eqnarray}}
\def\beq{\begin{equation}}
\def\eeq#1{\label{#1}\end{equation}}
\def\be{\begin{equation}}
\def\ee{\end{equation}}
\def\bea{\begin{eqnarray}}
\def\eea{\end{eqnarray}}

\def\s{s_\beta}
\def\c{c_\beta}
\def\mT{m_{{\rm T}_+}}
\def\pd{\partial}
\def\EPS2{\eps^2}
\def\half{\frac{1}{2}}
\def\r{x_t}
\def\lsim{\mathrel{\raise.3ex\hbox{$<$\kern-.75em\lower1ex\hbox{$\sim$}}}}
\def\gsim{\mathrel{\raise.3ex\hbox{$>$\kern-.75em\lower1ex\hbox{$\sim$}}}}
\def\ifmath#1{\relax\ifmmode #1\else $#1$\fi}

\def\maxack{\footnote{Work supported by the National Science
Foundation under grant PHY-0355005.}}

\begin{document}
\begin{titlepage}
\begin{flushright}
{\tt hep-ph/0506042} \\
\end{flushright}

\vskip.5cm
\begin{center}
{\huge \bf Electroweak Precision Constraints\CR on the Littlest
Higgs Model \CR  \vskip.3cm with T Parity} \vskip.2cm
\end{center}

\begin{center}
{\bf {Jay Hubisz, Patrick Meade, Andrew Noble, Maxim Perelstein\maxack}} \\
\end{center}
\vskip 8pt

\begin{center}
{\it Institute for High Energy Phenomenology, \\
F.R. Newman Laboratory of Elementary Particle Physics, \\
Cornell University, Ithaca, NY 14853, USA } \\
\vspace*{0.3cm}{\tt  hubisz@mail.lepp.cornell.edu,
meade@mail.lepp.cornell.edu\\ an76@cornell.edu,
maxim@mail.lepp.cornell.edu}
\end{center}

\vglue 0.3truecm

\begin{abstract}
\vskip 3pt \noindent
We compute the leading corrections to the properties of W and Z bosons
induced at the one-loop level in the SU(5)/SO(5) Littlest Higgs model
with T parity, and perform a global fit to precision electroweak data
to determine the constraints on the model parameters. We find that a
large part of the model parameter space is consistent with data.
Values of the symmetry breaking scale $f$ as low as 500 GeV are allowed, 
indicating that no significant fine tuning in the Higgs potential is 
required. We identify a region within the allowed parameter space in
which the lightest T-odd particle, the partner of the hypercharge gauge boson, 
has the correct relic abundance to play the role of dark matter. In 
addition, we find that a consistent fit to data can be obtained for 
large values of the Higgs mass, up to 800 GeV, due to the possibility of
a partial cancellation between the contributions to the $T$ parameter from
Higgs loops and new physics.

\end{abstract}

\end{titlepage}

\newpage

\renewcommand{\thefootnote}{\arabic{footnote}}
\setcounter{footnote}{0}

\section{Introduction}

The mechanism of electroweak symmetry breaking (EWSB) remains the
most pressing puzzle in elementary particle physics.
Experimentally, this question will be addressed at the Large
Hadron Collider (LHC). Theoretically, several interesting
possibilities have been proposed. In this article, we will
concentrate on the ``Little Higgs'' proposal~\cite{LH,LHreview}. In
this approach, the Higgs emerges as a pseudo-Nambu-Goldstone
boson, whose properties are constrained by global symmetries.
These global symmetries are not exact, and their breaking allows
the  Higgs to participate in non-derivative (i.e. gauge and
Yukawa) interactions. At the same time, there is enough global
symmetry left to ensure that the Higgs mass term vanishes at tree
level, and is only logarithmically sensitive to the unknown short
distance (ultraviolet, or UV) physics at the one-loop level. The usual 
quadratic
sensitivity of the Higgs mass parameter on the UV physics first
appears at two loops, and the incalculable UV effects remain
subleading as long as the cutoff of the theory (the scale at which
it becomes strongly coupled) is at or below about 10 TeV. With
this requirement, the Higgs mass terms are dominated by the
one-loop contribution from the top loops, which has the
appropriate sign to trigger the electroweak symmetry breaking, and
produces the Higgs vacuum expectation value (vev) of the right order of magnitude. Thus, Little Higgs theories provide an attractive explanation of EWSB.

The originally proposed implementations of the Little Higgs
approach suffered from severe constraints from precision
electroweak measurements~\cite{CC,HPR,PPP,CD}, which could only be
satisfied by finely tuning the model parameters. The most serious
constraints resulted from the tree-level corrections to precision
electroweak observables due to the exchanges of additional heavy
gauge bosons present in the theories, as well as from the small
but non-vanishing vev of an additional weak-triplet scalar field.
Motivated by these constraints, several new implementations of the
Little Higgs were proposed~\cite{simplest,custodial,ami}. Particularly
interesting is the approach of Refs.~\cite{LHT0,LHT,LHT1}, which introduces
a discrete symmetry, dubbed ``T parity'' in analogy to R parity
in the minimal supersymmetric standard model (MSSM). T parity
explicitly forbids any tree-level contribution from the heavy
gauge bosons to the observables involving only standard model (SM) particles 
as external states. It also forbids the interactions that induced the
triplet vev. As a result, in T parity symmetric Little Higgs
models, corrections to precision electroweak observables are
generated exclusively at loop level. This implies that the
constraints are generically weaker than in the tree-level case,
and fine tuning can be avoided~\cite{LHT0,LHT,LHT1}. 

The main goal of this paper is to investigate the electroweak precision
constraints on the models with T parity in more detail. We will concentrate
on the T parity symmetric version of the Littlest Higgs (LH) model, based
on an $SU(5)/SO(5)$ global symmetry breaking pattern~\cite{LHT}. Some 
phenomenological aspects of this model have been analyzed in Ref.~\cite{JP}.
The model possesses an attractive dark matter candidate, the T-odd partner
of the hypercharge gauge boson, which has the correct relic abundance in 
certain regions of the parameter space. It also leads to an interesting set 
of signatures at the LHC; in particular, an excess of events with large 
missing transverse energy is expected. In this paper, we will compute the 
corrections to the properties of $W/Z$ bosons induced by the new particles
present in the LH model, and perform a global fit to precision electroweak 
observables. We will show that a consistent fit can be obtained in a large 
region of the model parameter space, so that no significant fine tuning is 
required. We will also demonstrate that the LH model allows for consistent 
fits with values of the Higgs mass as large as 800 GeV, far in excess of the
upper bound obtained within the standard model. Finally, we will show that 
there exists a non-vanishing overlap between the region allowed by precision 
electroweak fits and the region where the LH model provides all of the 
observed dark matter.

The rest of the paper is organized as follows. After briefly
reviewing the $SU(5)/SO(5)$ Littlest Higgs model with T parity in
Section~\ref{model}, we will present the calculation of the corrections to 
precision electroweak observables in Section~\ref{electroweak}. In 
Section~\ref{fits},
we present the constraints on the parameter space of the model resulting from  
a global fit to precision electroweak observables. Section~\ref{conc} contains
our conclusions. A discussion of some aspects of the LH model in the 
renormalizable $R_\xi$ 
gauge, which we find useful in our calculations, is presented in the 
Appendix~\ref{app}.

\section{The Model}
\label{model}

In this section, we will review the LH model with T parity~\cite{LHT}, 
emphasizing the features that will be important for the analysis of this paper.

\subsection{Gauge-Scalar Sector}

The Littlest Higgs model~\cite{LH} embeds the electroweak sector of the 
standard model in an $SU(5)/SO(5)$ non-linear sigma model (nl$\sigma$m).  A 
global $SU(5)$ symmetry is broken to $SO(5)$ by the vev of an $SU(5)$ 
symmetric tensor $\Sigma$ of the form
\beq
\Sigma_0 \,=\, \left(\begin{array}{ccccc}
0& 0& 0& 1& 0\\
0& 0& 0& 0& 1\\
0& 0& 1& 0& 0\\
1& 0& 0& 0& 0\\
0& 1& 0& 0& 0\\
\end{array}\right).
\eeq{sigma0}
The low energy dynamics of the nl$\sigma$m is described in terms of  the 
field
\begin{equation}
\Sigma = e^{2 i \Pi/f} \Sigma_0,
\end{equation}
where $\Pi$ is the ``pion matrix'' containing the Goldstone degrees of 
freedom, and $f\sim1$~TeV is the nl$\sigma$m symmetry breaking scale, or 
``pion decay constant''. An $[SU(2)\times U(1)]^2$ subgroup of the global 
$SU(5)$ symmetry is gauged. The gauged generators have the form
\begin{eqnarray}\label{gauged}
&Q_1^a=\left( \begin{array}{ccc} \sigma^a/2 &0 & 0 \\
0 & 0 & 0\\ 0 & 0 & 0
\end{array}\right), \ \ \ &Y_1=
{\rm diag}(3,3,-2,-2,-2)/10\,,\nonumber \\
&Q_2^a=\left( \begin{array}{ccc} 0 & 0 & 0\\
0 & 0 & 0 \\
0 &0&-\sigma^{a*}/2\end{array} \right), & Y_2={\rm
diag}(2,2,2,-3,-3)/10~.
\end{eqnarray}
The kinetic term for the $\Sigma$ field can be written as
\beq
{\cal L}_{\rm kin} \,=\, \frac{f^2}{8} \mathrm{Tr} D_\mu \Sigma (D^\mu 
\Sigma)^\dagger,
\eeq{kinL}
where
\begin{equation}
D_\mu \Sigma =
\partial_\mu \Sigma - i \sum_j \left[ g_j W_j^a (Q_j^a \Sigma +
\Sigma Q_j^{aT} )+ g'_j B_j( Y_j \Sigma + \Sigma Y_j)\right]\,,
\end{equation}
with $j=1,2$.  Here, $B_j$ and $W^a_j$ are the $U(1)_j$ and
$SU(2)_j$ gauge fields, respectively, and $g^\prime_j$ and $g_j$ are the 
corresponding coupling constants. The vev $\Sigma_0$ breaks the extended 
gauge group $\left[ SU(2)
\times U(1) \right]^2$ down to the diagonal subgroup, which is identified 
with the standard model electroweak group
$SU(2)_L \times U(1)_Y$.  The fourteen pions of the $SU(5)/SO(5)$ breaking 
decompose into representations of the electroweak gauge group as follows:
\begin{equation}
\mathbf{1_0} \oplus \mathbf{3_0} \oplus \mathbf{2_{1/2}}
\oplus \mathbf{3_{\pm1}}.
\end{equation}
We will denote the fields in the above four representations as
$\eta$, $\omega$, $H$ and $\phi$, respectively. The field $H$ has the 
appropriate quantum numbers to be identified with the SM Higgs; after EWSB, 
it can be decomposed as $H=(-i\pi^+, \frac{v+h+i\pi^0}{\sqrt{2}})^T$, where 
$v=246$ GeV is the EWSB scale and $h$ is the physical Higgs field. Explicitly, 
the pion matrix in terms of these fields has the form
\beq
\Pi\,= \left(\begin{array}{ccccc}
-\omega^{0}/2- \eta/\sqrt{20} & -\omega^+/\sqrt{2} & -i \pi^+/\sqrt{2} & -i 
\phi^{++} & -i
\frac{\phi^{+}}{\sqrt{2}} \\
-\omega^-/\sqrt{2} & \omega^0/2- \eta/\sqrt{20} & \frac{v+h+i \pi^0}{2}& -i
\frac{\phi^{+}}{\sqrt{2}} & \frac{-i \phi^0 +\phi_P^0}{\sqrt{2}} \\
i \pi^-/\sqrt{2} & (v+h-i \pi^0)/2 & \sqrt{4/5} \eta  & -i
\pi^+/\sqrt{2} & (v+h+i \pi^0)/2 \\
i \phi^{--} & i \frac{\phi^{-}}{\sqrt{2}} & i \pi^-/\sqrt{2} &
-\omega^0/2 - \eta/\sqrt{20} &
- \omega^-/\sqrt{2} \\
i \frac{\phi^{-}}{\sqrt{2}} & \frac{i \phi^0 +\phi_P^0}{\sqrt{2}} &
\frac{v+h-i \pi^0}{2} & - \omega^+/\sqrt{2} &\omega^0/2- \eta/\sqrt{20}
\end{array}\right),
\eeq{pions}
where the superscripts indicate the electric charge. The fields $\eta$ and 
$\omega$ are eaten\footnote{In the LH model with T parity, the fields $\eta$ 
and $\omega$ mix with the field $\phi$ at order $(v/f)^2$, and it is a linear 
combination of these that is eaten. See Appendix~\ref{app} for details.} when 
the extended gauge group is broken down to $SU(2)_L\times U(1)_Y$, whereas 
the $\pi$ fields are absorbed by the standard model $W/Z$ bosons after  
EWSB. The fields $h$ and $\phi$ remain in the spectrum. Including the EWSB 
effects, the vev of the $\Sigma$ field has the form
\beq
\Sigma \,=\, \left(\begin{array}{ccccc}
0& 0& 0& 1& 0\\
0& -\half(1-c_v)& \frac{i}{\sqrt{2}}s_v& 0& \half(1+c_v)\\
0& \frac{i}{\sqrt{2}}s_v& c_v& 0& \frac{i}{\sqrt{2}}s_v\\
1& 0& 0& 0& 0\\
0& \half(1+c_v)& \frac{i}{\sqrt{2}}s_v& 0& -\half(1-c_v)\\
\end{array}\right),
\eeq{sigmafull}
where
\beq
s_v=\sin\frac{\sqrt{2}v}{f},~~~c_v=\cos\frac{\sqrt{2}v}{f}.
\eeq{sincos}
These formulas will prove very useful for analyzing the spectrum of the model.

The gauge generators are embedded in the $SU(5)$ is such a way that any given 
generator commutes with an $SU(3)$ subgroup of the $SU(5)$. This implies that 
if one pair of gauge couplings ($g_1, g_1^\prime$ or $g_2, g_2^\prime$) is 
set to zero, the Higgs field $H$ would be an exact Goldstone boson and, 
therefore, exactly massless.
Thus, any diagram renormalizing the Higgs mass vanishes unless it involves at 
least two of the gauge couplings. At one loop, all diagrams satisfying this 
property are only logarithmically divergent: the ``collective'' symmetry 
breaking mechanism protects the Higgs mass from quadratic divergences. The 
first quadratic divergence appears at two loop level.

The original Littlest Higgs model described above turned out to be 
significantly constrained by precision electroweak 
observables~\cite{CC,PPP,CD}.
T parity, a discrete $Z_2$ symmetry, was introduced by Cheng and 
Low~\cite{LHT0,LHT} to avoid this difficulty, and it also provides
a potential weak scale dark matter candidate. In the gauge sector,
T parity is an automorphism of the gauge groups which exchanges
the $\left[SU(2) \times U(1) \right]_1$ and $\left[SU(2) \times U(1) \right]_2$
gauge fields~\cite{LHT}.  The Lagrangian in Eq.~\leqn{kinL} is invariant under 
this transformation provided that
$g_1=g_2$ and $g^\prime_1=g^\prime_2$. In this case, the
gauge boson mass eigenstates (before EWSB) have the simple form, $W_\pm = 
(W_1 \pm W_2)/\sqrt{2}$, $B_\pm =  (B_1 \pm B_2)/\sqrt{2}$, where $W_+$ and 
$B_+$ are the standard model gauge bosons and are T-even, whereas $W_-$ and 
$B_-$ are the additional, heavy, T-odd states. (Typically, $B_-$ is the 
lightest T-odd state, and plays the role of dark matter~\cite{JP}.) After  
EWSB, the T-even neutral states $W_+^3$ and $B_+$ mix to produce the SM $Z$ 
and the photon. Since they do not mix with the heavy T-odd states, the 
Weinberg angle is given by the SM relation, $\tan\theta_w=g^\prime/g$, 
where $g=g_{1,2}/\sqrt{2}$ and $g^\prime=g_{1,2}^\prime/\sqrt{2}$ are the SM 
gauge couplings, and $\rho=1$ at tree level. As will be shown below, all the 
SM fermions are also T-even, implying that the $W_-$ and $B_-$ states generate
no corrections to precision electroweak observables at tree level.

The transformation properties of the gauge fields under T parity and the
structure of the Lagrangian~\leqn{kinL} imply that T parity acts on 
the pion matrix as follows: 
\beq
T: \Pi \rightarrow - \Omega \Pi \Omega\,,
\eeq{Tpi}
where $\Omega = \mathrm{diag}(1,1,-1,1,1)$.  This transformation
law ensures that the complex $SU(2)_L$ triplet $\phi$ is odd under
T parity, while the Higgs doublet $H$ is even. The trilinear coupling of the 
form $H^\dagger \phi H$ is therefore forbidden, and no triplet vev is 
generated. Eliminating this source of tree-level custodial $SU(2)$ violation 
further relaxes the precision electroweak constraints on the model.

\subsection{Light Fermion Sector}\label{light}

In the original LH model, the fermion sector of the standard model remained unchanged with the exception of the third generation of quarks, where the top Yukawa coupling had to be modified to avoid the large quadratically divergent contribution to the Higgs mass from top loops. In the model with T parity, however, the standard model fermion doublet spectrum needs to be doubled to avoid
compositeness constraints~\cite{LHT}. For each lepton/quark doublet, two fermion doublets $\psi_1\in (\mathbf{2,1})$
and $\psi_2\in (\mathbf{1,2})$ are introduced. (The quantum numbers refer to 
representations under the $SU(2)_1\times SU(2)_2$ gauge symmetry.) These can be embedded in incomplete representations $\Psi_1,\Psi_2$ of the global $SU(5)$ symmetry. An additional set of fermions forming an $SO(5)$ multiplet $\Psi^c$, transforming nonlinearly under the full $SU(5)$, is introduced to give mass to the extra fermions; the field content can be expressed as follows:
\begin{equation}
\begin{array}{ccc}
\Psi_1=\left(\begin{array}{c} \psi_1 \\ 0 \\ 0 \end{array}\right)\,,
& \Psi_2=\left(\begin{array}{c} 0 \\ 0 \\ \psi_2
\end{array}\right) \,,&
\Psi^c=\left(\begin{array}{c} \psi^c \\ \chi^c \\ \tilde{\psi}^c
\end{array}\right).
\end{array}
\end{equation}
These fields transform under the $SU(5)$ as follows:
\beq
\Psi_1 \rightarrow V^* \Psi_1\,, \hspace{.2in} \Psi_2 \rightarrow V
\Psi_2\,, \hspace{.2in}\Psi^c \rightarrow U\Psi^c,
\eeq{su5}
where $U$ is the nonlinear transformation matrix defined in 
Refs.~\cite{LHT,LHT1,JP}. The action of T parity on the multiplets takes 
\beq
\Psi_1\leftrightarrow -\Sigma_0 \Psi_2, ~~~\Psi^c \to -\Psi^c.
\eeq{tpar}
These assignments allow a term in the Lagrangian of the form
\begin{equation}\label{heavyyuk}
\kappa f (\bar{\Psi}_2 \xi \Psi^c+\bar{\Psi}_1 \Sigma_0 \Omega
\xi^\dagger \Omega \Psi^c),
\end{equation}
where $\xi=\exp(i\Pi/f)$. This term gives a Dirac mass $M_-=\sqrt{2}\kappa f$ to the T-odd linear combination of $\psi_1$ and $\psi_2$, $\psi_-=(\psi_1+\psi_2)/\sqrt{2}$, together with $\tilde{\psi}^c$; the T-even linear combination, $\psi_+=(\psi_1-\psi_2)/\sqrt{2}$, remains massless and is identified with the standard model lepton or quark doublet. To give Dirac masses to the remaining T-odd states
$\chi^c$ and $\psi^c$, additional fermions with opposite gauge
quantum numbers can be introduced~\cite{LHT,LHT1,JP}.

To complete the discussion of the fermion sector, we introduce the usual SM set of the $SU(2)_L$-singlet leptons and quarks, which are T-even and can participate in the SM Yukawa interactions with
$\psi_+$. The Yukawa interactions induce a one-loop quadratic divergence in the Higgs mass; however, the effect is numerically small except for the third generation of quarks. The Yukawa couplings of the third generation must be modified to incorporate the collective
symmetry breaking pattern; this is discussed in the next subsection.

\subsection{Top Sector} 

In order to avoid large one-loop quadratic divergences from the top sector, the $\Psi_1$ and
$\Psi_2$ multiplets for the third generation must be completed to
representations of the $SU(3)_1$ (``upper-left corner'') and $SU(3)_2$ 
(``lower-right corner'') subgroups of $SU(5)$.  These multiplets are
\begin{equation}
\begin{array}{ccc}
{\cal Q}_1=\left(\begin{array}{c} q_1 \\ U_{L1} \\ 0 \end{array}\right) \,,&
{\cal Q}_2=\left(\begin{array}{c} 0 \\ U_{L2} \\ q_2
\end{array}\right);
\end{array}
\end{equation}
they obey the same transformation laws under T parity and the $SU(5)$ symmetry as do $\Psi_1$ and $\Psi_2$, see Eqs.~\leqn{su5} and~\leqn{tpar}. The quark doublets are embedded such that
\begin{equation}
q_i = -\sigma_2 \left(\begin{array}{c} u_{Li} \\ b_{Li}
\end{array}\right).
\end{equation}
In addition to the SM right-handed top quark field $u_R$, which is assumed to be T-even, the model contains two $SU(2)_L$-singlet fermions $U_{R1}$ and $U_{R2}$ of hypercharge 2/3, which transform under T parity as
\begin{equation}
U_{R1}\leftrightarrow -U_{R2}.
\end{equation}
The top Yukawa couplings arise from the Lagrangian of the form
\beqa
\mathcal{L}_t &=& \frac{1}{2\sqrt{2}}\lambda_1 f \epsilon_{ijk}
\epsilon_{xy} \big[ (\bar{{\cal Q}}_1)_i \Sigma_{jx} \Sigma_{ky}  -
(\bar{{\cal Q}}_2 \Sigma_0)_i \tilde{\Sigma}_{jx} \tilde{\Sigma}_{ky}
\big] u_R \arline && \hspace{1in} + \lambda_2 f (\bar{U}_{L1}
U_{R1} + \bar{U}_{L2} U_{R2})+ {\rm h.c.}
\eeqa{topyuk}
where $\tSigma=\Sigma_0\Omega \Sigma^\dagger \Omega
\Sigma_0$ is the image of the $\Sigma$ field under T parity, see Eq.~\leqn{Tpi}, and the indices $i,j,k$ run from 1 to 3 whereas $x,y=4,5$. The T parity eigenstates are given by
\beq
q_\pm = \frac{1}{\sqrt{2}}(q_1 \mp q_2),~~~~
U_{L\pm} = \frac{1}{\sqrt{2}}(U_{L1} \mp U_{L2}),~~~~
U_{R\pm} = \frac{1}{\sqrt{2}}(U_{R1} \mp U_{R2}).
\eeq{rot1}
In terms of these eigenstates, Eq.~\leqn{topyuk} has the form
\beq
{\cal L}^T_{\rm m} =
\, \lambda_1 f \left[ \half(1+c_v) \bar{U}_{L+} + \frac{s_v}{\sqrt{2}} \bar{u}_{L+} \right] u_R
+ \lambda_2 f \left( \bar{U}_{L+} U_{R+} + \bar{U}_{L-} U_{R-} \right) +
{\rm h.c.}
\eeq{topmass}
where we have used Eq.~\leqn{sigmafull}. The T-odd
states $U_{L-}$ and $U_{R-}$ combine to form a Dirac fermion $T_-$,
with mass $m_{T_-}=\lambda_2 f$. The remaining T-odd states $q_-$ receive a Dirac mass from the interaction in Eq.~\leqn{heavyyuk}, and are assumed to be decoupled. The mass terms for the T-even states are diagonalized by defining
\beqa
t_L &=& \cos\beta \,u_{L+} - \sin\beta \,U_{L+},~~~~~~~
T_{L+} = \sin\beta \,u_{L+} +\cos\beta \,U_{L+},\CR
t_R &=& \cos\alpha \,u_R - \sin\alpha \,U_{R+},~~~~~~~
T_{R+} = \sin\alpha \,u_R + \cos\alpha \,U_{R+},
\eeqa{rot2}
where $t$ is identified with the SM top and $T_+$ is its T-even heavy partner. The mixing angles are given by
\beqa
\alpha &=& \half \tan^{-1} \frac{4\lambda_1\lambda_2 (1+c_v)}{4\lambda_2^2-\lambda_1^2(2s_v^2+(1+c_v)^2)}\, ,\CR
\beta &=& \half \tan^{-1} \frac{2\sqrt{2}\lambda_1^2 s_v (1+c_v)}{4\lambda_2^2+(1+c_v)^2\lambda_1^2-2\lambda_1^2 s_v}.
\eeqa{mixingsexact}
To leading order in the $v/f$ expansion,
\beq
\sin\alpha \,=\, \frac{\lambda_1}{\sqrt{\lambda_1^2+\lambda_2^2}},~~~
\sin\beta \,=\, \frac{\lambda_1^2}{\lambda_1^2+\lambda_2^2}\,\frac{v}{f}.
\eeq{mixings}
The masses of the two T-even Dirac fermions are given by
\beq
m^2_{t,T_+} \,=\,f^2\Delta\,\left(1\pm \sqrt{1-\frac{\lambda_1^2\lambda_2^2s_v^2}{2\Delta^2}}\right),
\eeq{massesexact}
where
\beq
\Delta \,=\, \half\,\left(\lambda_2^2+\frac{\lambda_1^2}{2}(s_v^2+\half(1+c_v)^2)
\right).
\eeq{Delta}
To leading order in $v/f$, 
\beq
m_t=\frac{ \lambda_1 \lambda_2
v}{\sqrt{\lambda_1^2+\lambda_2^2}},~~~~\mT = \sqrt{\lambda_1^2+\lambda_2^2}\, f.
\eeq{tmass}
 It is interesting to note that the T-odd states do not participate in the cancellation of quadratic divergences in the top sector: the cancellation only involves loops of $t$ and $T_+$, and the details are identical to the LH model without T parity~\cite{PPP}.

Using the above equations, it is straightforward to obtain the Feynman rules for the top sector of the LH model; we list the rules relevant for the calculations in this paper in Table~\ref{tab:Feynman}.

\begin{table}[t]
\center{\begin{tabular}{|c|c||c|c|}
\hline Particles & Vertices & Particles & Vertices \\
\hline
 $Z\bar{t}t$ & $\frac{e\c^2}{s_wc_w}\,\left( T^3 -\frac{2}{3}\frac{s_w^2}{\c^2} \right)$ & $Z\bar{T}_-T_- $ &  $-\frac{2}{3}\frac{es_w}{c_w}$\arline \hline
 $Z\bar{T}_+T_+ $ &$\frac{e}{s_wc_w}\,(\s^2 T^3 - \frac{2}{3}s_w^2 )$ & 
 $W \bar{b}_Lt_L $ & $\frac{e\c}{s_w}$ \arline \hline
 $Z\bar{t}T_+ $ & $-\frac{e\s\c}{s_wc_w}\,T^3$  
 & $W \bar{b}_L T_{+L} $ & $-\frac{e\s}{s_w}$ \CR
 \hline
\end{tabular}}
\caption{Feynman rules relevant for the analysis of the top sector of the LH model with T parity in this paper. We have defined $\s\equiv\sin\beta, \c\equiv\cos\beta$; $s_w, c_w$ denote the sine and cosine of the weak mixing angle; and $T^3=Q_1^3+Q_2^3$ is the diagonal generator of $SU(2)_L$.}\label{tab:Feynman}
\end{table}

\section{Corrections to Precision Electroweak Observables}
\label{electroweak}

The introduction of T parity automatically eliminates the tree level electroweak precision constraints that plagued the original Littlest Higgs model: since the external states in all experimentally tested processes are T-even, no T-odd state can contribute to such processes at tree level. The only non-SM T-even state in our model, the heavy top $T_+$, can only contribute at tree level to observables involving the SM top quark\footnote{Tree-level contributions of $T_+$ to 
other observables are 
suppressed by small off-diagonal CKM matrix elements.}, such as its couplings 
to $W$ and $Z$ bosons~\cite{Snow05}. At present, however,
these couplings have not been measured experimentally, so no contraints 
arise at the tree level.
At one-loop level, however, precision electroweak observables receive contributions from the $T_+$ 
as well as the T-odd particles. It is these contributions that determine the allowed parameter space of the Littlest Higgs model with T parity. We will evaluate the leading corrections in this section, and use them to perform a global fit to precision electroweak observables in Section~\ref{fits}.

In the SM, one-loop contributions to precision electroweak
observables from the top sector, enhanced by powers of the top
Yukawa coupling $\lambda_t$, dominate over contributions from the gauge and scalar sectors. We expect that the same hierarchy of effects
will hold in the Littlest Higgs model, and our main focus will be on analyzing the effects of the top sector. However, we will also include the custodial-symmetry violating contributions from the gauge sector and the T-odd partners of light fermions, which become important in certain regions of the parameter space. In addition, we will show explicitly that the contributions from the complex scalar triplet, which were shown to be potentially important in the original LH model~\cite{CD}, completely decouple in the T parity symmetric case due to the absence of the triplet vev.

Before proceeding with the calculations, let us make the following comment. The nl$\sigma$m underlying the LH model is a non-renormalizable effective theory, valid up to a cutoff scale $\Lambda\sim 4\pi f$. Every operator consistent with the symmetries of the low-energy theory will be generated at the cutoff scale $\Lambda$, and will contribute to the precision electroweak observables. However, we do not include such operators in the fit. (The only exception we make is to include the leading operator contributing to the $T$ parameter, Eq.~\leqn{oper}, since this parameter plays the most important role in constraining the model.) This is justified by the following considerations. First, while the contribution of the TeV-scale states that we will compute and the operator contributions that we will ignore are naively of the same order, $v^2/(16\pi^2 f^2)\sim v^2/\Lambda^2$, the former are logarithmically enhanced by a factor of $\log(f^2/v^2)\,\sim\,\log(\Lambda^2/f^2)\,\sim\,\log(4\pi)^2\,\sim 5$. Second, while a cancellation between the corrections computed below and the operator contribution is in principle possible if the UV physics produces an operator with a large coefficient, any significant change in the fits due to such a cancellation would represent a fine-tuning between the effects generated at two different energy scales, $f$ and $\Lambda$. 

\subsection{Oblique Corrections} \label{oblique}

\begin{figure}[t]
\centerline{\includegraphics[width=5in]{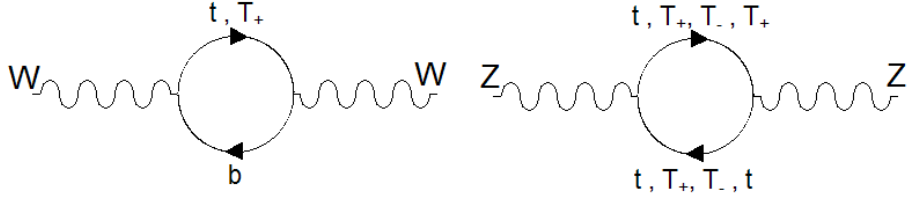}}
\caption{The one-loop diagrams contributing to the oblique corrections from
the top sector of the LH model with T parity.} \label{fig:STU}
\end{figure}

The largest corrections to precision electroweak observables in the LH model are induced by the one-loop diagrams involving the T-even $T_+$ quark shown in Fig.~\ref{fig:STU}. These oblique corrections can be described in terms of the Peskin-Takeuchi $S$, $T$, and $U$ parameters~\cite{peskintakeuchi}. The calculation of these parameters is straightforward if somewhat tedious; the result is
\beqa
S &=& \frac{\s^2}{2\pi}\,\left[ \left(\frac{1}{3}- \c^2\right)\,\log \r \,+\,\frac{(1+\r)^2}{(1-\r)^2}\,+\,\frac{2\r^2(3-\r)\log \r}{(1-\r)^3}\,-\frac{8}{3}
\right]\,, \CR
T &=& \frac{3}{16\pi}\,\frac{\s^2}{s_w^2c_w^2}\,\frac{m_t^2}{m_Z^2}\,\left[\frac{\s^2}{\r}-1-\c^2-\frac{2\c^2}{1-\r}\log \r \right]\,,\CR
U &=& -\frac{\s^2}{2\pi}\,\left[ \s^2\,\log \r \,+\,\frac{(1+\r)^2}{(1-\r)^2}\,+\,\frac{2\r^2(3-\r)\log \r}{(1-\r)^3}\,-\frac{8}{3}
\right] \,,
\eeqa{exactSTU}
where $\r=m_t^2/\mT^2$, $\s$ is the sine of the left-handed $t-T_+$ mixing angle given in Eq.~\leqn{mixingsexact}, and $s_w$ is the sine of the Weinberg angle. In the limit when $\r\ll1$, these formulas simplify considerably and we obtain
\beqa
S &=& \frac{1}{3 \pi} \left( \frac{\lambda_1}{\lambda_2} \right)^2
\frac{m_t^2}{\mT^2} \left[
  -\frac{5}{2} + \log \frac{\mT^2}{m_t^2} \right]\,, \arline
T &=& \frac{3}{8 \pi} \frac{1}{s_w^2c_w^2}\,\left(
\frac{\lambda_1}{\lambda_2} \right)^2\, \frac{m_t^4}{\mT^2 m_Z^2}
\,\left[ \log \frac{\mT^2}{m_t^2} - 1 + \frac{1}{2} \left(
\frac{\lambda_1}{\lambda_2} \right)^2 \right]\,, \arline U &=&
\frac{5}{6 \pi} \left( \frac{\lambda_1}{\lambda_2} \right)^2
\frac{m_t^2}{\mT^2}.
\eeqa{STUapprox}
The leading-order result for the $T$ parameter is in agreement with the previous analyses of LH models with and without T parity~\cite{CC,CD,LHT}. 

In contrast to $T_+$, the T-odd top partner $T_-$ does not
contribute to $S, T$ or $U$ since it is an $SU(2)_L$ singlet which does not mix with the SM top. However, $T_-$ loops do affect precision electroweak observables at the level of $(m_Z/m_{T_-})^2$ corrections which are not captured by the formalism of Peskin and Takeuchi. The corrections to the two most precisely measured observables, $s_W\equiv (1-m_W^2/m_Z^2)^{1/2}$ and $s_*$ (the value of the weak mixing angle implied from $Z$ decay asymmetries), are given by
\beq
s_*^2-s_0^2=-\frac{2\alpha}{45\pi} \frac{c_w^2 s_w^2}{c_w^2-s_w^2}\,\frac{m_Z^2}{m_{T_-}^2},~~~s_W^2-s_*^2=0,
\eeq{tminus}
where $s_0$ is the reference value of the weak mixing angle inferred from
\beq
s_0^2 (1-s_0^2) \,=\, \frac{\pi\alpha}{\sqrt{2}G_F m_Z^2}.
\eeq{sref}
These corrections are very small, and we do not include them in the global fit performed in Section~\ref{fits}.

It is clear from Eqs.~\leqn{STUapprox} that the $T$ parameter induced by the $T_+$ loops is about 20 times larger than the $S$ and the $U$ for the same model parameters, and therefore the constraints on the model are largely driven by the $T$ parameter. This parameter also receives a contribution from the gauge sector of the model. While in general subdominant, this correction becomes important when the $t-T_+$ mixing is suppressed (namely when the ratio $\lambda_1/\lambda_2$ is small), and we will include it in the fit. This contribution arises from the custodial $SU(2)$-violating tree level mass splitting of the T-odd heavy $W^3_H$ and $W^{\pm}_H$ gauge bosons. Neglecting effects of order $g^{\prime2}$, the mass splitting is given by
\beq
\Delta M^2 \equiv M^2(W^3_H)-M^2(W^\pm_H) = \frac{g^2 f^2}{4}\left(1-c_v\right)^2 \approx \frac{1}{8} g^2 \frac{v^4}{f^2}.
\eeq{deltaM}
At one loop, this effect induces a contribution to the $T$ parameter~\cite{LHT}:\footnote{We find that the calculation of this contribution simplifies considerably in the Landau gauge, $\xi=0$.}
\beq
T_{W_H} = - \frac{9}{16 \pi c_w^2 s_w^2 M_Z^2} \Delta M^2 \log
  \frac{\Lambda^2}{M^2(W_H)}.
\eeq{Tgauge1}
Note that the result is divergent, and depends on the UV cutoff of the theory $\Lambda$. 
This should not be surprising since the theory we're dealing with is non-renormalizable: indeed, the mass splitting in Eq.~\leqn{deltaM} comes from a dimension-6 operator of the form $W^\mu W_\mu (H^\dagger H)^2$, which appears when the $\Sigma$ fields in Eq.~\leqn{kinL} are expanded to order $\Pi^4$. The UV divergence signals the presence of a ``counterterm'' operator of the form~\cite{LHT}
\beq
{\cal L}_c \,=\,\delta_c \frac{g^2}{16\pi^2}\,f^2\,\sum_{i,a} {\rm Tr}\,\left[
(Q_i^a D_\mu \Sigma)(Q_i^a D^\mu \Sigma)^*\right]\,,
\eeq{oper}
where $\delta_c$ is an order-one coefficient whose exact value depends on the details of the UV physics. (The normalization of Eq.~\leqn{oper} is fixed by naive dimensional analysis~\cite{NDA}.) Including the counterterm, the full contribution of the gauge sector to the $T$ parameter has the form
\beq
T\,=\, -\frac{1}{4\pi s_w^2}\,\frac{v^2}{f^2} \left( \delta_c+\frac{9}{8} \log\frac{\Lambda^2}{M^2(W_H)} \right)\,=\, -\frac{1}{4\pi s_w^2}\,\frac{v^2}{f^2} \left( \delta_c+\frac{9}{4}\log\frac{4\pi}{g} \right).
\eeq{Tgauge}
where we have assumed $\Lambda=4\pi f$ and used $M(W_H)=gf$. 
As expected, this contribution is parametrically subdominant to the correction from the top sector, Eq.~\leqn{STUapprox}, in the limit $m_t\gg m_Z$. Furthermore, in agreement with the discussion at the beginning of this section, the effect of the operator~\leqn{oper} induced at the cutoff scale is subdominant, by a factor of $1/\log(\Lambda^2/M^2(W_H))\sim 0.2$, compared to the calculable contribution in Eq.~\leqn{Tgauge1}. 

In Section~\ref{fits}, we will also be interested in the effects of varying the Higgs mass. To leading order, the Higgs contribution to the oblique parameters is given by~\cite{peskintakeuchi}
\beqa
S &=& \frac{1}{12 \pi} \log
\frac{m_h^2}{m_{h,\mathrm{ref}}^2}\,, \arline 
T &=&-\frac{3}{16 \pi c_w^2} \log \frac{m_h^2}{m_{h,\mathrm{ref}}^2}\,,
\arline U &=& 0,
\eeqa{STUhiggs}
where $m_{h,\rm{ref}}\ll m_h$ is the ``reference'' value of the Higgs mass used to obtain the SM predictions for precision electroweak observables\footnote{It should be kept in mind that in the LH model, the Higgs couplings to the $W/Z$ bosons will receive corrections of order $v/f$, which have been neglected in Eq.~\leqn{STUhiggs}. This will not affect any of the conclusions of our analysis in
Section~\ref{fits}.}.
Interestingly, the negative contribution to the $T$ parameter from a heavy Higgs can be partially cancelled by the positive contribution from the  $T_+$. As we will show below, this cancellation allows for a consistent fit to precision electroweak observables with the Higgs mass well above the upper bound obtained in the SM~\cite{PDG}. 

Finally, the LH model contains an additional T-odd $SU(2)_L$-triplet scalar field $\phi$, with the mass
\beq
m_\phi^2\,\approx\,\frac{2m_h^2 f^2}{v^2}\,\sim (1 {\rm~TeV})^2.
\eeq{mphi}
After EWSB, a mass splitting of order $v^2/f$ between various components of the triplet is generated, for example, by operators of the form $H^\dagger \phi \phi^\dagger H$. Neglecting this mass splitting, the triplet contributions to $S$, $T$ and $U$ parameters vanish; keeping the terms of order $(m_Z/m_\phi)^2$, its contributions to $s_*$ and $s_W$ are given by
\beq
 s_*^2-s_0^2\,=\,-\frac{\alpha}{24\pi}\frac{s_w^2 c_w^2}{c_w^2-s_w^2}\,\frac{m_Z^2}{m_\phi^2}\,,~~~s_W^2-s_*^2\,=\,\frac{\alpha}{60\pi}\frac{m_W^2}{m_\phi^2}.
\eeq{tripobs}
If the mass splitting is taken into account, non-zero contributions to the Peskin-Takeuchi parameters are induced; however, these effects are of order $\Delta m_\phi^2/m_\phi^2 \sim v^4/m_\phi^4$, and are thus subleading to the corrections given in Eq.~\leqn{tripobs}. We conclude that the effects of the triplet $\phi$ on the precision electroweak observables in the LH model with T parity decouple with growing $m_\phi$, and are negligible for $m_\phi$ in its natural range, around 1 TeV. We will not include these effects in the global fit of Section~\ref{fits}. 

\subsection{Effects of the T-Odd Partners of Light Fermions}
\label{stuff}

\begin{figure}[t]
\centerline{\includegraphics[width=4in]{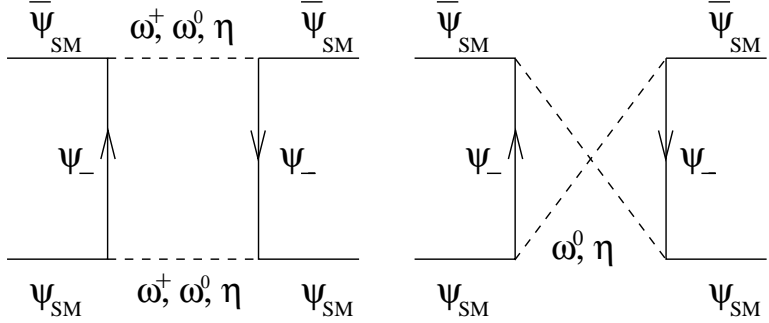}} 
\caption{The box diagrams which provide the leading contribution to the four-fermion operators in the limit $\kappa \gg g$.}
\label{baksho}
\end{figure}

To implement T parity in the fermion sector of the LH model, it is necessary to introduce a T-odd fermion partner for each lepton/quark doublet of the SM (see Section~\ref{light}). These particles are vector-like, and their effects on precision electroweak observables must decouple in the limit when their mass is taken to infinity. However, box diagrams involving the exchanges of Goldstone bosons $\omega$ and $\eta$, see Fig.~\ref{baksho}, generate four-fermion operators whose coefficients {\it increase} if the mass of the T-odd fermions is increased while $f$ is kept fixed\footnote{We are grateful to Thomas Gregoire for bringing this point to our attention.}. This non-decoupling is easy to understand qualitatively: to increase the mass of the T-odd fermions, it is necessary to increase the Yukawa coupling $\kappa$ in Eq.~\leqn{heavyyuk}, which in turn makes the four interaction vertices in the box diagrams stronger. Assuming that the couplings $\kappa$ are flavor-diagonal and flavor-independent, the generated operators have the form
\beq
{\cal O}_{\rm 4-f}=  - \frac{\kappa^2}{128 \pi^2 f^2}\,\bar{\psi}_L
\gamma^\mu \psi_L \bar{\psi}^\prime_L \gamma_\mu \psi^\prime_L\,,
\eeq{4fermi}
where $\psi$ and $\psi^\prime$ are (distinct) SM fermions, and we ignore the 
corrections of order $g/\kappa$. The experimental bounds on four-fermi interactions involving SM fields provide an {\it upper} bound on the T-odd fermion masses; the strongest constraint comes from the $eedd$ operator, whose coefficient is required to be smaller than $2\pi/(26.4~{\rm TeV})^2$~\cite{PDG}. This yields
\beq
M_{\rm TeV} \lsim 4.8 f^2_{\rm TeV}\,,
\eeq{upperbound}
where $M_{\rm TeV}$ and $f_{\rm TeV}$ are the values of the T-odd fermion masses and the symmetry breaking scale, respectively, in TeV. In addition, a lower
bound on the masses of the T-odd fermions can be obtained from non-observation 
of these particles at the Tevatron; in analogy with squarks of the MSSM, we
expect the bound to be in the neighborhood of 250--300 GeV.

The contribution of each T-odd doublet to the $T$ parameter is given by
\beq
T_{\rm T-odd}= -\, \frac{\kappa^2}{192 \pi^2 \alpha} \left(\frac{v}{f}\right)^2 \,,
\eeq{ST}
where we omit terms of order $(v/f)^4$ and higher. Note that, for a fixed value of $f$, this contribution {\it increases} with increasing T-odd fermion mass; Eq.~\leqn{upperbound} implies that 
\beq
|T_{\rm T-odd}|\lsim 0.05, 
\eeq{upperT}
independent of $f$. (Note that this bound relies on the assumption that 
the $\kappa$ couplings are flavor-independent.)  
Nevertheless, the T-odd fermions can have a noticeable effect on the precision electroweak fits due to a large number (twelve) of doublets in the SM; this 
will be illustrated in the next section. 

\if
With the assumptions of flavor-diagonal Yukawa couplings $\kappa$, the non-oblique corrections induced by the T-odd fermions at one loop are flavor-universal, and can be absorbed in the $S$ parameter. 
\draftnote{[The $T$ parameter contribution at order $(v/f)^4$ is divergent. as is the $S$ parameter already at $(v/f)^2$. Including those contributions would require adding counterterms, so the number of parameters in the fit would increase making the fits more complicated. This is beyond the scope of this paper.]} 
\fi

\subsection{$Z\to b\bar{b}$ Vertex Renormalization}
\label{zbbar}

\begin{figure}[t]
\centerline{\includegraphics[width=5in]{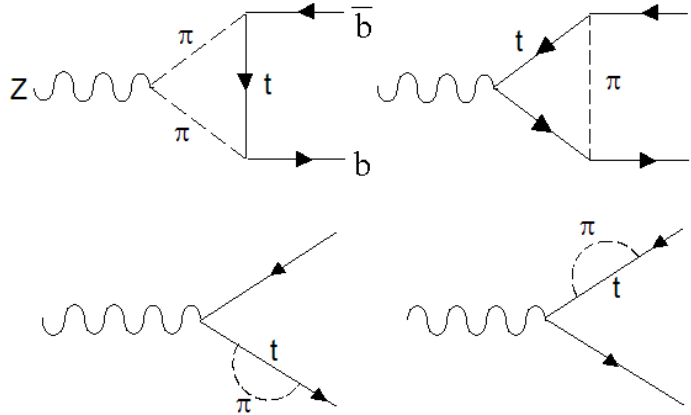}} 
\caption{The dominant standard model diagrams which contribute to the 
$Zb\bar{b}$ vertex renormalization at one-loop in the $R_\xi$ gauge.}
\label{zbbsmloops}
\end{figure}

In the SM, the largest non-oblique correction is the renormalization of the  $Z b \bar{b}$ vertex by top quark loops. This effect is non-decoupling in the sense that it is proportional to the square of the top mass. This non-decoupling is most easily seen if the calculation is performed in the 't Hooft-Feynman gauge~\cite{pich}. In this gauge, the non-decoupling part of the vertex correction comes purely from the diagrams involving the exchange of a Goldstone boson $\pi^\pm$, since its couplings to the top and bottom quarks are enhanced by the top Yukawa $\lambda_t$. These diagrams are shown in Fig.~\ref{zbbsmloops}. The diagrams involving the exchange of the gauge bosons are subdominant in the large-$m_t$ limit, and neglecting their contribution only induces an error of order $(m_Z/m_t)^2\sim 25$\% in the vertex correction calculation.  

\begin{figure}[t]
\centerline{\includegraphics[width=4in]{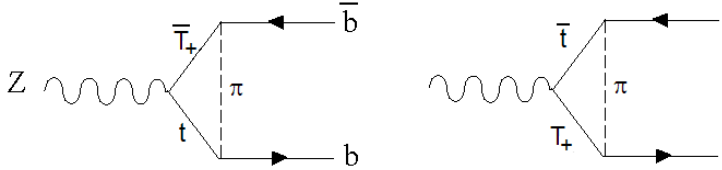}} 
\caption{The additional ``mixing'' diagrams contributing to the $Zb\bar{b}$
vertex renormalization at one loop in the LH model with T parity.}
\label{zbbbsmloops}
\end{figure}

We have calculated the one-loop correction to the $Zb\bar{b}$ vertex in the LH model with T parity. We used the 't Hooft-Feynman gauge (for a brief discussion of the $R_\xi$ gauges in the LH model, see Appendix~\ref{app}). As in the SM case, the diagrams involving $\pi^\pm$ exchanges dominate in the large-$m_t$ limit, and we neglected all other contributions. (While a more precise calculation could be done, the effort would not be justified as the $Zb\bar{b}$ correction turns out to have only a small effect on precision electroweak fits.) These diagrams are of three kinds. First, the same diagrams as in the SM appear, but with the top coupling to the $Z$ 
modified according to Table~\ref{tab:Feynman}. Second, all the diagrams in Fig.~\ref{zbbsmloops} also appear with the top replaced by the T-even heavy top partner $T_+$. Third, the ``mixing'' diagrams shown in Fig.~\ref{zbbbsmloops} appear
as a result of the mixing between $t$ and $T_+$. These diagrams can be easily calculated using the couplings given in Table~\ref{tab:Feynman} and in Eq.~\leqn{GBcouple} of Appendix~\ref{app}. To leading order in the limit $\mT\gg m_t \gg m_W$, the result is
\beq
\delta g_L^{b\bar{b}}=\frac{g}{c_w}\frac{\alpha}{8\pi s_w^2} \frac{m_t^4}{m_W^2\mT^2}
\frac{\lambda_1^2}{\lambda_2^2} \log \frac{\mT^2}{m_t^2},
\eeq{zbb}
where $\delta g_L^{b\bar{b}}$ is the correction received by the $Zb_L\bar{b}_L$ vertex in the LH model {\it in addition} to the usual SM one-loop correction. It is interesting to note that this leading order contribution comes entirely from the mixing diagrams in Fig.~\ref{zbbbsmloops}. The correction to the $Zb_R\bar{b}_R$ vertex is negligible since it is not enhanced by the top Yukawa coupling.
Note also that the correction in Eq.~\leqn{zbb} does not have the correct sign 
to alleviate the well-known deviation of the measured value of the 
forward-backward asymmetry in $Z\to b\bar{b}$ decays from the SM 
prediction  inferred from the other precision electroweak 
observables~\cite{Chan}. 

A calculation of $Zb\bar{b}$ in a general theory containing an extra heavy quark that mixes with $t$ has been carried out in
Ref.~\cite{burgessrb}.  Accounting for the fact that $U_+$ is a
vector isosinglet and including the appropriate mixing specific to
the LH model, Eq.~\leqn{zbb} agrees with the results of this analysis in the limit $\mT\gg m_t \gg m_W$. The results of Ref.~\cite{burgessrb} are more general, valid for arbitrary values of $\mT$ and $m_t$. However, we find that using these expressions instead of Eq.~\leqn{zbb} does not lead to noticeable changes in the global fits performed in Section~\ref{fits}.

With the assumption of flavor-diagonal and flavor-independent Yukawa couplings
$\kappa$ made in Section~\ref{stuff}, the one-loop vertex corrections due to 
loops of T-odd fermions are flavor-universal, and can therefore be absorbed 
in the redefinitions of gauge couplings. They will not induce an observable
shift in $Z b\bar{b}$ couplings.


\section{Constraints on the Littlest Higgs Parameter Space}
\label{fits}

\begin{figure}[tb]
\centerline{\includegraphics[width=0.5\hsize]{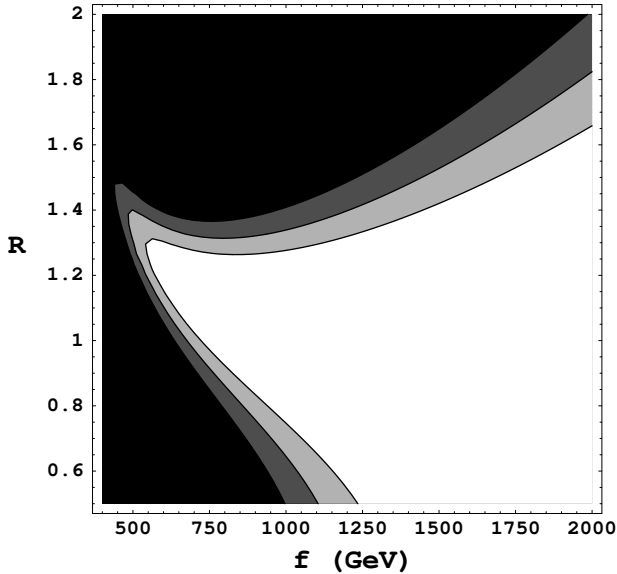}}
\vskip-0.3cm
\caption{Exclusion contours in terms of the parameter $R=\lambda_1/\lambda_2$ 
and the symmetry breaking scale $f$. The contribution of the T-odd fermions 
to the T parameter is neglected. From lightest to darkest, the contours 
correspond to the 95, 99, and 99.9 confidence level exclusion.
} 
\label{fR}
\end{figure}

\begin{figure}[tb]
\centerline{\includegraphics[width=0.5\hsize]{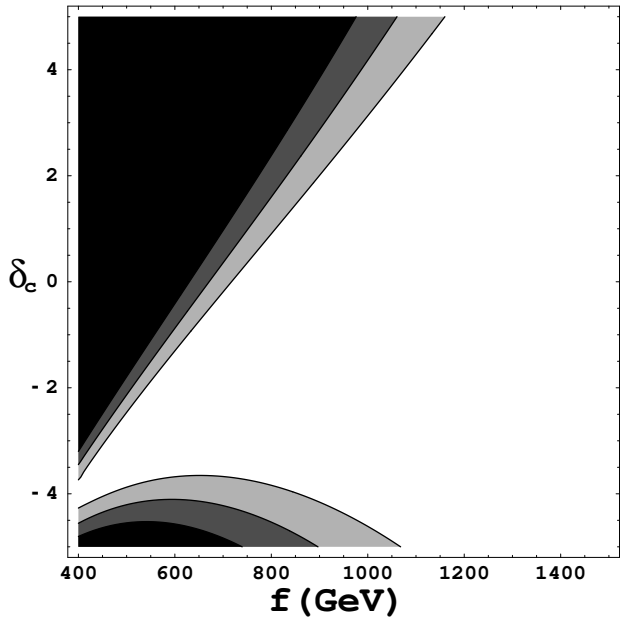}}
\vskip-0.3cm
\caption{Exclusion contours in terms of the UV contribution to custodial 
symmetry violation $\delta_c$,  see Eq.~\leqn{oper}, and $f$. 
From lightest to darkest, the contours correspond to the 95, 99, and 99.9 
confidence level exclusion.}
\label{deltac}
\end{figure}

To obtain constraints on the parameter space of the LH model with T parity, we have performed a global fit to precision electroweak observables, including the LH contributions evaluated in the previous section. The LH contributions are parametrized by two dimensionless numbers, $R=\lambda_1/\lambda_2$ and $\delta_c$, and the symmetry breaking scale $f$. In the fit, we have used the values of the  21 $Z$ pole and low-energy observables listed in Ref.~\cite{PDG}; the equations expressing the shifts in these observables in terms of the oblique parameters and $\delta g^{b\bar{b}}_L$ are given in Ref.~\cite{burgess1}. We take the top mass to be 176.9 GeV~\cite{PDG}, and do not include the uncertainty associated with the top mass. In each constraint plot, we draw the $95$, $99$, and $99.9\%$
confidence level contours in the context of a $\chi^2$ analysis with
two degrees of freedom\footnote{It is important to note that changing the 
assumed number of degrees of freedom can strongly affect the positions of the
contours; this is equivalent to modifying the priors that enter into
the fit~\cite{strumiaandguido}. A complete Bayesian analysis taking into
account a variety of different priors for the model parameters is beyond
the scope of this paper, but it would be straightforward to perform such
an analysis using the formulas provided in Section~\ref{electroweak}.}.
                                                                               
In the first part of the analysis, we have fixed the Higgs mass at its 
reference value, $m_{h,{\rm ref}}=113$ GeV. In Fig.~\ref{fR}, we plot the 
constraints in the $f-R$ plane, assuming $\delta_c=0$. 
In Fig.~\ref{deltac}, we fix $R=1$ and plot the constraints in the $f-\delta_c$ plane, neglecting the T-odd fermion contribution. It is clear that a large part of the parameter space is consistent with precision electroweak constraints, including regions where the symmetry breaking scale $f$ is as low as 500 GeV. (In these regions, a partial cancellation between the top and gauge sector contributions to the $T$ parameter takes place.) In some cases, we have even obtained consistent fits for values of $f$ as low as 
350 GeV. However, since our analysis neglects all higher-derivative operators generated at the scale $\Lambda\sim4\pi f$, which can contribute significantly to precision electroweak observables when $\Lambda\lsim5$ TeV, we estimate that the fits cannot be trusted for $f\lsim400$ GeV, and do not show that part of the parameter space in the plots.
 
\begin{figure}[t]
\centerline{\includegraphics[width=0.5\hsize]{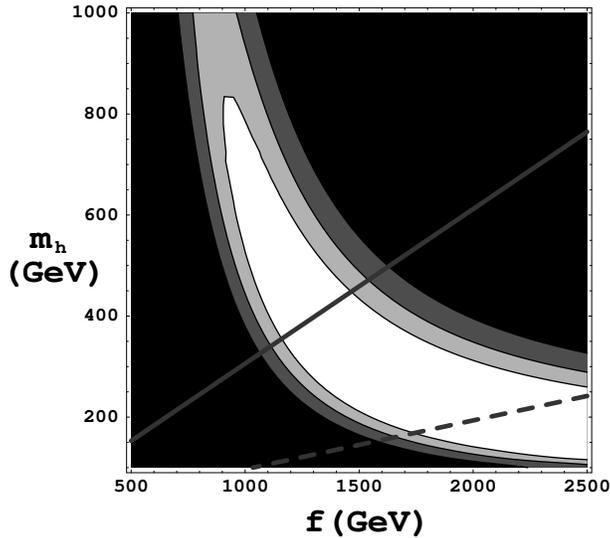}}
\vskip-0.3cm
\caption{Exclusion contours in terms of the Higgs mass $m_h$ and the symmetry breaking scale $f$. From lightest to darkest, the contours correspond to the 95, 99, and 99.9 confidence level exclusion. Contours of constant values of the fine-tuning parameter $F$ are also shown; the solid and dashed lines correspond to $F=10$ and $F=100$, respectively.}
\label{hmass}
\end{figure}

As has been shown in Section~\ref{oblique}, top sector loops in the LH model provide a sizable, positive contribution to the $T$ parameter. This raises an interesting possibility: since the contribution of a heavy SM Higgs to the $T$ parameter is negative, it is possible that these two effects partially cancel\footnote{A consistent fit with a heavy Higgs can also be obtained in the Littlest Higgs model {\it without} T parity, where a positive correction to the $T$ parameter is generated at tree level; however, this requires a rather high value of $f$, of order $5$ TeV~\cite{strumiaandguido}. A similar cancellation of the heavy Higgs and new physics contributions to $T$ also occurs in top seesaw models~\cite{topseesaw}; see Ref.~\cite{Sekhar}. We are grateful to Bogdan Dobrescu for bringing this paper to our attention.}, and a consistent fit is obtained for $m_h$ far in excess of the usual SM upper bound, currently about 250 GeV.
This possibility is illustrated in Fig.~\ref{hmass}, where we fix $R=2$, $\delta_c=0$, and plot the constraints in the $f-m_h$ plane. Remarkably, values of $m_h$ as high as 800 GeV are allowed at 95\% confidence level. (Note that the approximation made in Eq.~\leqn{STUhiggs}, where the corrections of order $v/f$ in the Higgs contribution to the oblique parameters have been neglected, is justified in the region of interest, since $f$ is still of order 1 TeV.) Thus, the LH model provides an explicit, well-motivated example of a theory in which the SM upper bound on the Higgs mass is avoided. 
Moreover, from the point of view of fine tuning in the Higgs potential, the high values of $m_h$ are more natural in the context of this model~\cite{espinosa}. For example, let us use the ratio of the one-loop top contribution to $m_h^2$ to the full $m_h^2$,
\beq
F \,=\, \frac{3\lambda_t^2 \mT^2}{4\pi^2m_h^2}\,\log\frac{\Lambda^2}{\mT^2},
\eeq{finetune}
as a quantitative measure of the fine tuning. (Larger values of $F$ correspond to higher degree of fine tuning.) Plotting the contours of constant $F$ indicates that, in the region of the parameter space consistent with precision electroweak constraints, the degree of fine tuning increases with the decreasing Higgs mass. Large values of $m_h$ are clearly preferred from the point of view of naturalness in the Higgs potential.

\begin{figure}[t]
\centerline{\includegraphics[width=0.5\hsize]{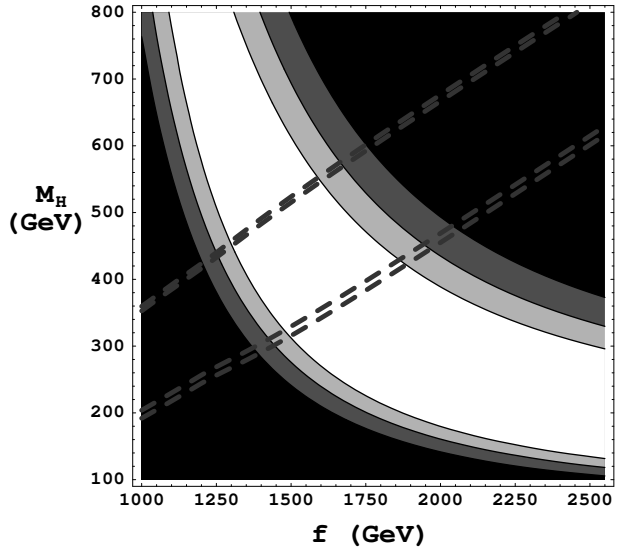}}
\vskip-0.3cm
\caption{In this plot, the lines of constant relic density of the LTP
are superimposed upon the constraints from the precision 
electroweak observables in the $f-m_h$ plane. In the narrow bands between the 
pairs of dashed lines, the LTP relic density is within 2$\sigma$ 
of the central value provided by the WMAP collaboration~\cite{WMAP}. For the 
detailed analysis of the LTP relic density, see Ref.~\cite{JP}.
} \label{darkmattertoo}
\end{figure}

If T parity is an exact symmetry (including the theory 
completing the description above the scale $\Lambda$), the lightest T-odd particle (LTP) is stable. Generically, the 
LTP is the T-odd partner of the hypercharge gauge boson, which is electrically neutral and can play the role of WIMP dark matter. The
LTP relic density has been computed in Ref.~\cite{JP}, and a region in the parameter space where the LTP can account for all of the observed dark matter has been identified. In Fig.~\ref{darkmattertoo}, the 2 sigma contours on the dark matter relic density are superposed over a plot of the precision electroweak 
constraints where $R=2$, $\delta_c=0$, and 
$m_h$ and $f$ are allowed to vary. There is a region of the allowed parameter 
space in which the LTP can account for all of the dark matter.

\begin{figure}[tb]
\centerline{\includegraphics[width=0.5\hsize]{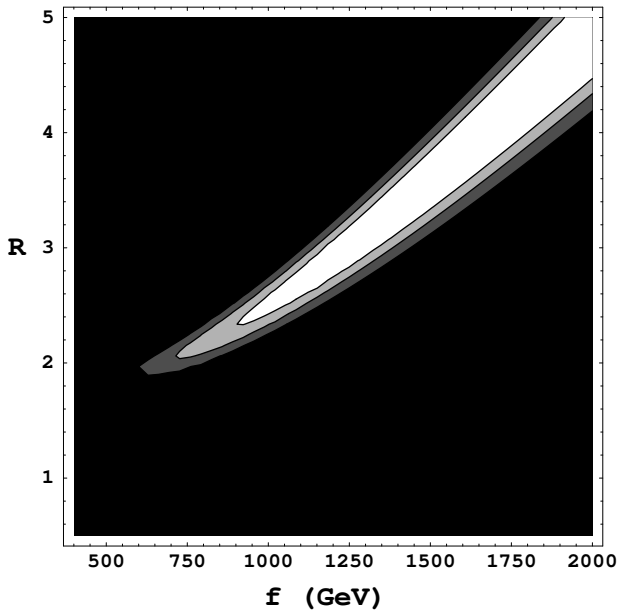}}
\vskip-0.3cm
\caption{Exclusion contours in terms of the parameter $R=\lambda_1/\lambda_2$ 
and the symmetry breaking scale $f$. The contribution of the T-odd fermions 
to the T parameter is included assuming that it
has the maximal size consistent with the constraint from four-fermi
interactions, Eq.~\leqn{upperT}. From lightest to darkest, the contours 
correspond to the 95, 99, and 99.9 confidence level exclusion.
} 
\label{fR1}
\end{figure}

In 
Figs.~\ref{fR}--\ref{darkmattertoo}, the 
contribution of the T-odd fermions to the $T$ parameter is neglected. This 
approximation is justified as long as the T-odd fermions
are sufficiently light: for example, for T-odd fermion mass of 300 GeV, their 
total contribution to the $T$ parameter is very small, and does not have
any noticeable effect on the fits. On the other hand, heavier T-odd fermions 
can have a substantial effect. This is illustrated in 
Fig.~\ref{fR1}, where the T-odd fermion contribution has been
assumed to have the maximal size consistent with the constraint from four-fermi
interactions, Eq.~\leqn{upperT}. (This corresponds to the T-odd fermion masses
saturating the upper bound in Eq.~\leqn{upperbound}.) While the constraints 
in this 
case are more severe, consistent fits can still be obtained for
values of $f$ below 1 TeV. 

\section{Conclusions}
\label{conc}

In this paper we have calculated the dominant corrections to the  
precision electroweak observables at the one-loop level in the Littlest Higgs 
model with
T parity~\cite{LHT}. We performed a global fit to the precision
electroweak observables and found that a large part of the model parameter 
space is consistent with data. In particular, a consistent fit can be obtained
for values of the nl$\sigma$m symmetry breaking scale $f$ as low as 500 
GeV. Furthermore, we found that the LH model can fit the data for values of
the Higgs mass far in excess of the SM upper bound, due to the possibility of
a partial cancellation between the contributions to the $T$ parameter from
Higgs loops and new physics. Combining our results with those of 
Ref.~\cite{JP}, we found 
that there are regions of parameter space allowed
by precision electroweak constraints where the lightest T-odd particle 
can account for all of the observed dark matter.

We have argued that the corrections to low energy observables in the LH model 
are dominated by the top sector, and our analysis was primarily focused on 
those contributions. It would be interesting to perform a more detailed 
analysis of the effects from the gauge and scalar sectors; however, we do not 
expect these effects to substantially modify our conclusions. The analysis
of the T-odd fermion sector in this paper relied on rather restrictive
simplifying assumptions: in particular, the Yukawa couplings in the T-odd
sector were assumed to be flavor-diagonal and flavor-independent. 
A possible non-trivial flavor structure of their couplings could have 
interesting experimental consequences. Moreover, these fermions should be 
sufficiently light to be pair-produced at the LHC, or even at the Tevatron.
It is therefore important to analyze that sector of the model in more detail.  

\if
These additional fermions are not required for naturalness, but are needed 
to implement the SM fermions in a way consistent with T parity and 
compositeness bounds. While their effects on the precision electroweak 
observables are expected to be small, a possible non-trivial flavor structure
of their couplings could have interesting experimental consequences. 
Moreover, these fermions could be sufficiently light to be pair-produced at
the LHC, and may affect the relevant phenomenology. It would be interesting to
analyze that sector of the model in more detail.  
\fi

In conclusion we find that the Littlest Higgs model with T parity is
only weakly constrained by precision electroweak data, and
provides a viable alternative for physics at the TeV scale. Apart from being 
theoretically attractive, the model has 
several features that are of interest for planning future experiments. Two 
examples are the possibility of a relatively heavy Higgs, as discussed
in this paper, and the similarity of many of the collider signatures of
this model to the benchmark SUSY signatures, which will inevitably complicate 
the LHC analysis~\cite{JP}. We hope that our analysis, which explicitly 
demonstrates the viability of the LH model, will open the door
for further detailed studies of its collider phenomenology.

\vskip.5cm
\section*{Acknowledgements}

We would like to thank Bogdan Dobrescu and Ian Low for useful discussions 
related to this work. We would also like to acknowledge helpful correspondence with Thomas Gregoire. This work is supported by the National Science
Foundation under grant PHY-0355005.

\vskip1cm

\appendix

\section{Renormalizable Gauges for the Littlest Higgs with T Parity}
\label{app}

While the higher-order corrections for observable quantities in
gauge theories must be gauge independent, an appropriate gauge
choice can greatly reduce the complexity of a loop calculation,
and make the underlying physics more transparent. This is
especially important in the case of spontaneously broken gauge
symmetries. While many issues in these theories are most easily
understood in the unitary gauge, this gauge is ill-suited for loop
calculations, leading to complicated intermediate expressions,
and, in some cases, ambiguous answers\footnote{A well-known example
of such an ambiguity appears in the calculation of the $W$ boson
contribution to the anomalous magnetic moment of the muon~\cite{Rxi}.}. 
Experience
 with the SM radiative correction calculations indicates that it 
is best to use the renormalizable, or $R_\xi$, gauges; a special case
of $\xi=1$, the 't~Hooft-Feynman gauge, is especially useful. In
this paper, we have used this gauge to calculate the $Zb\bar{b}$ coupling shift, see Section \ref{zbbar}. Since only unitary-gauge Feynman rules have appeared in the literature so far for the LH models~\cite{BPP,Han}, we will briefly discuss the  $R_\xi$ gauges for the LH model with T parity in this Appendix. We will focus on the charged gauge boson sector; the analysis of the neutral sector is similar. Even though the calculations in the paper do not require it, in this Appendix we will keep all correction of order $(v/f)^2$, since an interesting effect of $\omega-\phi$ mixing first appears at that order. 

The charged gauge boson mass matrix follows from Eq.~\leqn{kinL}; to order $\eps^2$ (where $\eps=v/f$) it has the form
\begin{equation}
\label{mass}
M^2 =
f^2\,\left(\begin{array}{cc}
                                       g_1^2& -g_1g_2(1-\eps^2/4) \\
                                       -g_1g_2(1-\eps^2/4)  & g_2^2
                                       \end{array}\right).
\end{equation} 
Diagonalizing this matrix results in the mass eigenstates
\begin{equation}
W_L = c_0 W_1 + s_0 W_2,~~~W_H =
-s_0 W_1 + c_0 W_2.
\end{equation}
In the LH model with T parity, the gauge couplings are
set equal, $g_1=g_2 = \sqrt{2} g$ and the mixing angle is given by $s_0 = c_0 = 1/\sqrt{2}$.
The charged gauge boson mass eigenvalues are then
\begin{equation}
M_H^2 = g^2 f^2 \left[ 1 - \frac{1}{4} \eps^2
 + \cdots \right], \ \ \ \ M_L^2 = \frac{g^2 v^2}{4} \left[ 1 - \frac{1}{6}
  \eps^2 + \cdots \right].
\end{equation}
Spontaneous breaking of the gauge symmetries leads to the kinetic mixing between the gauge bosons and the Goldstone boson fields in Eq.~\leqn{pions}. In the mass eigenbasis, the mixing terms have the form
\begin{eqnarray}
{\cal L}_{W \pi} &=& M_H W_H^{\mu -} \left[ \pd_\mu \o^+ \left(1 -
  \frac{1}{24} \eps^2 \right) - \frac{i}{6} \eps^2 \pd_\mu \phi^+ \right] \arline
&& + M_L W_L^- \pd_\mu \pi^+ \left( 1 - \frac{1}{12} \eps^2 \right)+ \mathrm{h.c.}
\end{eqnarray}
At order $\eps^2$, the Goldstone boson fields in Eq.~\leqn{pions} are not canonically normalized. To canonically normalize the Goldstone fields, we perform the following rescaling:
\begin{eqnarray}
\pi^\pm &\rightarrow& \pi^\pm \left( 1 + \frac{1}{12} \EPS2 \right)\,,
\arline
\o^\pm &\rightarrow& \o^\pm \left( 1 + \frac{1}{24} \EPS2 \right)\,,
\arline
\phi^\pm &\rightarrow& \phi^\pm \left( 1 + \frac{1}{24} \EPS2 \right).
\end{eqnarray}
After this redefinition, there are still kinetic mixing terms
involving the $\o$ fields and the complex triplet, $\phi$:
\begin{equation}
{\cal L}_\mathrm{kin} = \pd_\mu \o^+ \pd^\mu \o^- + \pd_\mu \phi^+
\pd^\mu \phi^- + \frac{i}{12} \EPS2 ( \pd_\mu \o^+ \pd^\mu \phi^- -
\pd_\mu \phi^+ \pd^\mu \o^- )\,.
\end{equation}
These terms are diagonalized with the redefinition
\begin{eqnarray}
\o'^\pm = \o^\pm \mp \frac{i}{24} \EPS2 \phi^\pm \,,\arline
\phi'^\pm = \phi^\pm \pm \frac{i}{24} \EPS2 \o^\pm\,.
\end{eqnarray}
In terms of these new canonically normalized fields, the gauge
boson-Goldstone mixing terms are given by\footnote{Note that the normalizations of the fields in the definition of the Goldstone boson matrix, Eq.~\leqn{pions}, have been chosen so that the mixing term has the simple form in Eq.~\leqn{kinmix}.}
\beqa
{\cal L}_{W \pi} &=& M_H W_H^{\mu -} \left[ \pd_\mu \o'^+ - \frac{i}{8} \EPS2 \pd_\mu \phi'^+ \right] \arline
&& + M_L W_L^- \pd_\mu \pi^+ + \mathrm{h.c.}
\eeqa{kinmix}
A final rotation which leaves the kinetic terms diagonal,
\begin{eqnarray}
\o''^\pm = \o'^\pm \mp \frac{i}{8} \EPS2 \phi'^\pm \,,\arline
\phi''^\pm = \phi'^\pm \mp \frac{i}{8} \EPS2 \o'^\pm\,,
\end{eqnarray}
identifies $\o''^\pm$ as the combination of Goldstones eaten by the
heavy gauge bosons, and $\phi''^\pm$ as the uneaten combination.  
A similar, but algebraically more involved, analysis carries through for the massive neutral gauge bosons.

Following the usual logic of renormalizable gauges, we add a gauge-fixing term which, after integration by parts, cancels the mixing terms:
\begin{equation}
\Delta {\cal L} =
\frac{1}{2 \xi_L} \left| \pd_\mu W_{L\mu}^\pm + M_L \xi_L \pi^\pm
\right|^2+\frac{1}{2 \xi_H} \left| \pd_\mu W_H^\pm + M_H \xi_H \o''^\pm \right|^2.
\end{equation}
The mass eigenstates in the eaten Goldstone sector are $\pi^\pm$, with mass $\sqrt{\xi_L}M_L$, and $\o''^\pm$, with mass $\sqrt{\xi_H} M_H$. (We have used the 't Hooft--Feynman gauge, $\xi_L=1$, in Section~\ref{zbbar}.) Note that the $\pi$ fields do not mix with $\o$ and $\phi$ at any order in $v/f$, since such mixing is forbidden by T parity. The situation would be considerably more involved in the case of the Littlest Higgs without T parity.

Given the exact identification of the $\pi^\pm$ fields in Eq.~\leqn{pions} with the lighter mass eigenstate in the $R_\xi$ gauges, it is straightforward to obtain the $bt\pi$ and $bT\pi$ vertices required in the calculation of Section~\ref{zbbar}. Expanding the $\Sigma$ matrices in Eq.~\leqn{topyuk} to linear order in $\Pi$, and using Eqs.~\leqn{rot1},~\leqn{rot2} to transform to the mass eigenbasis for the top sector, we obtain
\beq
-i\sqrt{2}\lambda_1 \bar{b}_{L+} \pi^- \, (\cos\alpha \,t_R + \sin\alpha\,T_{R+}) + {\rm h.c.} \,=\,-i\lambda_t \bar{b}_{L+} \pi^- \, \left(t_R + \frac{\lambda_1}{\lambda_2} T_{R+} \right)+ {\rm h.c.}
\eeq{GBcouple}
where $\lambda_t$ is the SM Yukawa coupling, and $b_+$ is identified with 
the SM $b$ quark. Note that the couplings involving the $b_+$ and any one 
of the T-odd Goldstone bosons, $\phi$, $\omega$ or $\eta$, vanish due to the 
structure of the Lagrangian~\leqn{topyuk} and the fact that the field $u_R$ 
is T-even.

\end{document}